\documentstyle{elsart}
\begin{document}
\begin{frontmatter}
\title{ Homogeneous nucleation: Comparison between two theories }
\author{Larissa V. Bravina$^{a,b,c,1}$, Eugene E. 
Zabrodin$^{a,b,c}$}
\address{
$^a$Department of Physics, University of Bergen,
All\`egaten 55, \\
N-5007 Bergen, Norway \\
$^b$Institute for Nuclear Physics, Moscow State University,
119899 Moscow, Russia\\
$^c$Institute for Theoretical Physics, University of Frankfurt,\\
Robert-Mayer-Str. 8-10, D-60054 Frankfurt, Germany \\
$^1$ Alexander von Humboldt Foundation Fellow}

\maketitle

\begin{abstract}
The classical nucleation theory of Becker, D\"{o}ring and Zeldovich
is compared with the Langer coarse-grained field approach to the 
nucleation phenomenon. Both formalisms have been applied to the 
condensation from a supersaturated vapor. It is shown that the
nucleation rate derived in the classical theory can be expressed
in a form equivalent to that of the field nucleation theory.
This equivalence serves as an explanation of the puzzling fact
that the numerical predictions of both theories for condensation
of Xe and CO$_2$ are almost identical though the standard 
analytical expressions for the nucleation rates are different.
The results obtained can help to link the theories of nucleation
and their approximations.\\


\end{abstract}
\end{frontmatter}

\newpage

First order phase transitions, which include the thoroughly studied 
condensation of a vapor, boiling of a liquid or crystallization of a 
melt, as well as the hypothetical transition from a hadronic phase
to the quark$-$gluon plasma, play an important role in modern science
and technology.
The theory of first order phase transitions is based on the 
homogeneous nucleation theory which aims to determine the rate of the
decay of a metastable state, i.e. the rate of formation of 
nucleating clusters within the initially homogeneous nonequilibrium
state. This is the first important assumption of the nucleation 
theory which implies that the conversion of a metastable state into
a thermodynamically stable phase should proceed via the formation of 
clusters of the new phase. As a typical example one may consider the
condensation of small drops of liquid from the vapor rapidly quenched
into the metastable region.

Clusters of various sizes are produced spontaneously due to the 
fluctuations in a homogeneous medium. Clusters with sizes larger than
a certain size, called critical, will grow into a new phase, and 
thus the phase transition process develops. A description of this 
process requires a kinetic analysis of the evolution of the clusters.
The phenomenological nucleation theory, often referred to as the
classical or Becker$-$D\"{o}ring$-$Zeldovich (BDZ) theory, was 
formulated in the pioneering works by Becker and D\"{o}ring 
\cite{BeDo35}, and Zeldovich \cite{Zeld42}. It is explained in detail 
in many textbooks and reviews, see e.g. Refs. 
\cite{Fren46,Zett69,Abr74,LiPi81,GMS83,Kelt91,PeLe79}.
Despite the phenomenological origin, the classical theory appears
to predict nucleation phenomena very well.

A field theoretical approach to the problem of nucleation has been 
proposed by Langer \cite{Lang67,Lang69,LaTu73} who extended the
generalization of the BDZ theory for a system of arbitrary many
degrees of freedom \cite{LaSw61} to the field theories. This 
formalism has been applied to several nucleation phenomena, 
including the liquid$-$vapor phase transition \cite{LaTu73}.
In spite of all the statistical and hydrodynamic corrections, it 
appears that the nucleation rate calculated within the coarse-grained
field theory differs only slightly from that of the BDZ theory, while
the expressions for the rates are different. This surprisingly good
agreement between the two theories leads to the assumption that, 
although the coarse-grained field theory is a more accurate and 
advanced treatment of the nucleation phenomena than the classical 
theory, the latter compensates somehow correctly for all degrees of 
freedom \cite{GMS83,Kelt91,LaTu73}.

In the present paper we would like to show that the nucleation rate 
derived in the BDZ theory may be obtained under certain assumptions
in the form which is identical to that of the Langer theory. We will 
start from a brief review of the classical nucleation theory. Then, 
after discussion of the formalism of the coarse-grained field theory, 
the results of both models will be applied to condensation from
a supersaturated vapor. The comparison will be done in terms of the 
Langer theory.

The metastable state typically is assumed to be a mixture of molecules 
and clusters of the new phase containing $n$ molecules. Let $N_n(t)$ 
be the average number of clusters of size $n$ at time $t$.  Effects of 
the direct interactions between the clusters are disregarded. 
Therefore, the clusters may change their size only due to an 
evaporation$-$condensation mechanism, in which a cluster of size $n$ 
grows or shrinks by condensation or loss of a single molecule. This is 
the basic assumption of the classical nucleation theory. The rate of 
change of the number of clusters of a given size is determined by the
master equation
\begin{equation}
\displaystyle
\frac{\partial N_n(t)}{\partial t} = J_n - J_{n-1}\ ,\ n \geq 2\ ,
\label{1}
\end{equation}
where 
\begin{equation}
\displaystyle
J_n = g_{n-1}^+\, N_{n-1}(t) - g_n^-\, N_n(t)
\label{2}
\end{equation}
is a current in a size space, i.e. it is the rate at which clusters of 
size $n-1$ grow to clusters of size $n$. Here $g_{n-1}^+$ is the rate
of the molecule condensation to a cluster of size $n$, and $g_n^-$ is
the rate of evaporation. According to the theory of thermodynamic 
fluctuations the equilibrium distribution of clusters obeys the
Boltzmann (or Gibbs) distribution
\begin{equation}
\displaystyle
N_n^{eq} = N_0^{eq} \exp \left(- \frac{\Delta G_n}{k_B T} \right)\ ,
\label{3}
\end{equation}
where $\Delta G_n$ is the minimum work needed to form a cluster of size
$n$, $k_B$ is Boltzmann's constant, $T$ is the temperature, and 
$N_0^{eq}$ is a pre-exponential factor. A maximum in $\Delta G_n$ is 
reached at $\partial \Delta G_n/ \partial n = 0$ corresponding to a
cluster of critical size, $n_{c}$. Clusters smaller than the critical
cluster tend to shrink, clusters of the critical size are in a 
metastable equilibrium, and clusters larger than the critical cluster 
tend to increase their volume. 

The equilibrium cluster distribution function $N_i^{eq}$ is a time 
independent solution of the master equation (\ref{1}). The current 
$J_n$ is zero for this distribution, therefore, one may write the
equation of detailed balance
\begin{equation}
\displaystyle
g_{n-1}^+ \exp \left(- \frac{\Delta G_{n-1}}{k_B T} \right) =
g_n^- \exp \left(- \frac{\Delta G_n}{k_B T} \right)\ .
\label{4}
\end{equation}
As shown by Zeldovich \cite{Zeld42}, under the treatment of $n$
as a continuous variable the master equation (\ref{1}) may be 
approximated by the Fokker$-$Planck kinetic equation
\begin{equation}
\displaystyle
\frac{\partial N_n(t)}{\partial t} = -\frac{\partial J_n}{\partial n}
= \frac{\partial }{\partial n} \left[ \frac{g_n}{k_B T} \frac{\partial
\Delta G_n}{\partial n} N_n(t) + g_n \frac{\partial N_n(t)}{\partial 
n} \right]\ , 
\label{5}
\end{equation}
or, in terms of the ratio $N_n(t)/\!N_n^{eq}$,
\begin{equation}
\displaystyle
\frac{\partial N_n(t)}{\partial t} = \frac{\partial }{\partial n}
\left[ g_n N_n^{eq} \frac{\partial}{\partial n} \left( \frac{N_n(t)}
{N_n^{eq}} \right) \right]\ .
\label{6}
\end{equation}

The classical theory is based on a nonequilibrium, steady-state 
solution of Eq.~(\ref{5}), corresponding to a continuous phase 
transition. Then $J_n = s = constant$, which is called the nucleation 
rate and measures the number of clusters passing through the critical
range per unit time per unit volume. 

The boundary conditions for small and large $n$ are chosen in the 
BDZ theory as follows,
\begin{equation}
\displaystyle
\lim_{n \rightarrow 0} \frac{N_n(t)}{N_n^{eq}}  = 1 \ \ \ ,
\ \ \ \lim_{n \rightarrow \infty} \frac{N_n(t)}{N_n^{eq}} = 0 \ .
\label{7}
\end{equation}
A number of small clusters is assumed to be close to its equilibrium
value due to the rapid increase of the fluctuation probability with 
decrease of cluster size. It is assumed also that clusters larger than
that of critical size are removed from the system.
The steady-state solution of Eqs.~(\ref{6}) and (\ref{7}) reads
\begin{equation}
\displaystyle
\frac{N_n(t)}{N_n^{eq}} = s \, \int_n^\infty \, 
\frac{d n^\prime}{g_n^\prime\, N_{n^\prime}^{eq}}\ ,
\label{8}
\end{equation}
where
\begin{equation}
\displaystyle
s^{-1} \, =\, \int_0^\infty \, 
\frac{d n^\prime}{g_n^\prime\, N_{n^\prime}^{eq}}\ .
\label{9}
\end{equation}
Since the function 
$ N_n^{eq} = N_0^{eq} \exp (- \Delta G_n / k_B T )$
has a very sharp maximum at $n_{c}$, the integral in (\ref{9}) can
be evaluated by the saddle-point method, i.e. by expanding of 
$\Delta G_n$ around $n_{c}$ and extending the integration with 
respect to $n - n_{c}$ from $-\infty$ to $\infty$. With these 
approximations, the result for the steady-state nucleation rate
may be written in the general form
\begin{equation}
\displaystyle
s = g_{n_{c}} N_{n_{c}}^{eq} Z\ ,
\label{10}
\end{equation}
containing the Zeldovich factor
\begin{equation}
\displaystyle
Z = \left( \frac{\gamma}{2 \pi k_B T} \right)^{1/2}\ \ \ ,
\ \ \ \gamma = - \left( \frac{\partial^2 \Delta G_n}{\partial n^2} 
\right)_{n = n_{c}}\ .
\label{11}
\end{equation}
Here the number of clusters that have passed through the critical 
range is expressed via the equilibrium number of clusters of critical
size. This is the principal result of the classical theory.

The accretion rate of the molecule condensation to the surface of
the critical cluster $g_{n_{c}}$ is derived from the kinetic theory
of gases to be
\begin{equation}
\displaystyle
g_{n_{c}} = \frac{p\, S_c}{\sqrt{2 \pi k_B T m_1}}\ ,
\label{11a}
\end{equation}
containing the gas pressure $p$, the surface of the critical droplet
$S_c$, and the average mass of a single molecule $m_1$. On the other
hand $g_{n_{c}}$ may be obtained from the solution of rather general 
equations describing the diffusion growth of dense clusters. We will 
come to this point later during the discussion of vapor condensation.

An extension of the classical theory to a system of arbitrary many
degrees of freedom was worked out by Landauer and Swanson 
\cite{LaSw61}. Their method was applied by Langer to the field 
nucleation theory \cite{Lang69}. The BDZ and the Langer theories 
postulate that unstable fluctuations of thermal origin, e.g. 
clusters of critical
size, lead to the decay of the metastable state. However, in the
field nucleation theory the critical cluster needs not necessarily be
a physical object \cite{CaHi59}, but rather characterizes a certain
saddle-point configuration. In contrast to the classical theory, only
the critical cluster and its shape deformations are required for the
field theory of nucleation. The advantages and shortcomings of both
models are discussed in detail in \cite{GMS83}.

The Langer theory starts with the consideration of a classical 
system with $\cal N$ degrees of freedom (coordinates and momenta)
$\eta_i,\ i = 1, \ldots , \cal N$. The probability distribution 
functional $\rho(\{\eta_i\},t)$ is assumed to obey a continuity
equation in $\{\eta\}$-space, namely
\begin{equation}
\displaystyle
\frac{\partial \rho(\{\eta_i\},t)}{\partial t} = 
-\sum_{i=1}^{\cal N} \frac{\partial J_i }{\partial \eta_i } \ ,
\label{12}
\end{equation}
which is analogous to the multi-dimensional Fokker$-$Planck equation.
The probability current $J_i$ is given by
\begin{equation}
\displaystyle
J_i = -\sum_{i=1}^{\cal N} M_{ij} \left( 
\frac{\partial F }{\partial \eta_i } \rho + k_B T
\frac{\partial \rho }{\partial \eta_i } \right) \ , 
\label{13}
\end{equation}
containing a generalized mobility matrix $M_{ij}$, and a 
coarse-grained free energy functional $F\{\eta\}$.
For the case of $J_i = 0$ the equilibrium solution of Eq.~(\ref{12})
reads 
\begin{equation}
\displaystyle
\rho_{eq}\{\eta_i\} \propto \exp \left(- \frac{F\{\eta_i\}}{k_B T} 
\right)\ .
\label{14}
\end{equation}
The basic idea of the field nucleation theory is that the states of 
metastable and stable equilibrium lie in the vicinity of the local 
minima of $F\{\eta_i\}$ in the $\{\eta\}$-space, which maximize 
$\rho_{eq}\{\eta_i\}$. The phase transition occurs when the 
configuration $\{\eta_i\}$ moves from the vicinity of a metastable
minimum to the vicinity of a stable minimum. To develop into the 
stable state the configuration must overcome the potential barrier. 
It is most likely for 
the trajectory of the system to pass across a small area around the 
intermediate saddle point of the functional $F$, say $\{\bar\eta\}$.
This saddle-point configuration $\{\bar\eta\}$ corresponds to
the critical cluster of a condensing phase in the BDZ theory,
although it may not be a physical object. The rate of the decay of
the metastable state is determined by the steady-state solution of
Eqs.~(\ref{12}) and (\ref{13}), corresponding to a finite probability
current across the saddle point from the metastable to the stable
minimum of $F$.

To obtain this solution Langer \cite{Lang69} introduced a set of new
variables $\xi_k,\ k=1, \ldots, \cal N$, whose fluctuations are slow
compared to the characteristic time of the evaporation$-$condensation
process. These variables should contain the order parameter. They are
related to the $\eta$'s by virtue of the orthogonal transformation 
$\hat D$:
\begin{equation}
\displaystyle
\xi_j = \sum_i D_{ij} \left( \eta_i - \bar \eta_i \right)\ ,
\label{15}
\end{equation}
where, near the saddle point $\{\xi\} = \{0\}$,
\begin{equation}
\displaystyle
F\{\eta\} = \bar F + {1 \over 2} \sum_{j=1}^{\cal N} \lambda_j 
\xi_j^2 + \, \ldots
\label{16}
\end{equation}
Here $\bar F = F\{\bar \eta\}$, and $\lambda_j$ are the eigenvalues of
the matrix $M_{ij} = ( \partial^2 F / \partial \eta_i \partial i
\eta_j )_{\eta = \bar \eta}$. By the definition of the saddle point at 
least one of the eigenvalues $\lambda_j$ must be negative. Following 
Langer we denote this eigenvalue as $\lambda_1$. 
Then in terms terms of the ratio 
$\sigma(\{\xi\}) = \rho(\{\xi\})/\!\rho_{eq}(\{\xi\})$
the requirement of a steady state current over the potential barrier 
has the form 
\begin{equation}
\displaystyle
\sum_{i}\frac{\partial J_i}{\partial \xi_i} = -\sum_{ij}M_{ij}\,
\frac{\partial}{\partial \xi_i}\, \left( \rho_{eq}(\{\xi\})
\frac{\partial \sigma(\{\xi\})}{\partial \xi_j} \right) = 0\ ,
\label{16a}
\end{equation}
which is a straightforward generalization of Eq.~(\ref{6}). The
stationary probability distribution functional $\rho(\{\xi\})$
obeys the boundary conditions that, near the metastable state,
$\rho(\{\xi\})$ is very close to the equilibrium distribution
$\rho_{eq}\{\eta_i\}$, and beyond the saddle point $\rho(\{\xi\})$
should vanish rapidly. The solution of Eq.~(\ref{16a}) which 
satisfies these boundary conditions becomes
\begin{equation}
\displaystyle
\sigma(\{\xi\}) = \frac{1}{\sqrt{2 \pi k_B T}}\,
\int_{u}^{\infty} \exp{\left( - \frac{a^2}{2 k_B T} \right)}\, da\ ,
\label{16b}
\end{equation}
where $u$ is a linear combination of the $\xi$'s
\begin{equation}
\displaystyle
u = \sum_{i}\, U_i \xi_i \ .
\label{16c}
\end{equation}
The coefficients $U_i$ are solutions of the eigenvalue equation
\begin{equation}
\displaystyle
\lambda_n\, \sum_{i}\, M_{ni} U_i = -\kappa U_i\ ,
\label{16d}
\end{equation}
which describe the growth rate $\kappa$ of the single unstable mode
at the saddle point. To obtain the desired nucleation rate one has 
to integrate the flux $J_i$ over any plane containing the saddle
point and not parallel to $J_i$. After the cumbersome but 
straightforward calculations the steady state solution of 
Eq.~(\ref{16a}) may be written in the form \cite{Lang69}
\begin{equation}
\displaystyle
I = I_0\, \exp \left( - \frac{\Delta F_{c}}{k_B T} \right)\ ,
\label{17}
\end{equation}
where 
\begin{equation}
\displaystyle
I_0= |\kappa|\, {\cal V}\, \left( \frac{k_B T}{2 \pi |\lambda_1|} 
\right)^{1/2}\, = \frac{|\kappa|}{2 \pi}\, \Omega_0\ .
\label{18}
\end{equation}
Here $\Delta F_{c}= \bar F(\bar \eta) - \bar F(\eta_0)$ is the excess
free energy required to form a cluster of the critical size. The
dynamical prefactor $\kappa$ describes the cluster growth at the 
saddle point, the statistical prefactor $\Omega_0$ is often called 
the generalization of the Zeldovich
factor, and $\cal V$ denotes the volume of the saddle point subspace.
Although Eq.~(\ref{17}) has a structure similar to Eq.~(\ref{10}), 
it is asserted that the prefactors in both equations should be 
different \cite{LaTu73}. However, up to now there is no crucial 
experiment which can distinguish the predictions of Langer theory from 
those of the BDZ theory. 

To compare the models let us consider the condensation of a 
supersaturated vapor. The minimum work needed to form a spherical 
droplet of radius $R$ in the thin wall approximation is simply
\begin{equation}
\displaystyle
\Delta F(R) = - \frac{4\pi}{3}R^3\Delta p \, +\, 4\pi R^2 \sigma \  ,
\label{19}
\end{equation}
where $\Delta p$ is the difference between the pressures inside and
outside the droplet and $\sigma$ is the surface tension. The critical
radius is given by the Laplace formula $R_{c} = 2 \sigma /\!\Delta 
p $, and near this value $\Delta F(R)$ can be represented by harmonic
approximation
\begin{equation}
\displaystyle
\Delta F(R)=\frac{4}{3}\pi \sigma R^2_c\, -\, 4 \pi \sigma 
(R -R_c)^2 \ .
\label{20}
\end{equation}
It is convenient to rewrite the excess of the free energy in terms of
the reduced radius $r = R/\!R_{c}$ and the new similarity number
$\displaystyle \lambda_Z = R_c \sqrt{ \frac{4 \pi \sigma}{k_B T} } $ 
introduced by the authors \cite{BrZa95} recently,
\begin{equation}
\displaystyle
\Delta F(R) = \frac{k_B T}{3} \lambda_Z^2 - k_B T (r - 1)^2 
\lambda_Z^2
\label{21}
\end{equation}
From the definition (\ref{16}) it follows that $\xi = r - 1$, the 
free energy of the critically large droplet is $\bar F = k_B T 
\lambda_Z^2 / 3$ and the only negative eigenvalue 
$\lambda_1$ is $\lambda_1 = 2 k_B T \lambda_Z^2$. Therefore, the 
nucleation rate predicted by the Langer theory becomes
\begin{equation}
\displaystyle
I_0^L = \frac{|\kappa|}{2 \sqrt{\pi}}\, \frac{{\cal V}}
{\lambda_Z}\ .
\label{22}
\end{equation} 
The statistical prefactor $\Omega_0$ is obviously determined by the
formula
\begin{equation}
\displaystyle
\Omega_0 = {\cal V}\, \left( \frac{2 \pi k_B T}{|\lambda_1|} \right)
 = {\cal V}\, \frac{\sqrt{\pi}}{\lambda_Z}\ .
\label{22a}
\end{equation}
Substituting Eq.~(\ref{21}) in Eq.~(\ref{11}) we get for the 
Zeldovich factor
\begin{equation}
\displaystyle
\gamma = - \left( \frac{\partial^2 \Delta F}{\partial R^2}\right)_{R_c}
 = 2 k_B T \left( \frac{\lambda_Z}{R_c} \right)^2
\label{22b}
\end{equation}
and 
\begin{equation}
\displaystyle
Z = \left( \frac{\gamma}{2 \pi k_B T} \right)^{1/2} = 
\frac{\lambda_Z}{\sqrt{\pi} R_c}\ .
\label{22c}
\end{equation}
The factor $Z$ has the physical meaning of a measure of the width 
$\Delta$ of the Gaussian distribution by which the actual distribution 
$\rho^{-1}(R)$ is approximated, namely $Z = \Delta^{-1}$. Therefore,
the statistical prefactor $\Omega_0$, which measures the volume of the
saddle point in the phase space, can not be considered as a simple
generalization of the Zeldovich factor. Instead, it is related to the 
$Z$ factor by
\begin{equation}
\displaystyle
\Omega_0 = \frac{{\cal V}}{R_c}\, Z^{-1}\ .
\label{22d}
\end{equation}

The nucleation rate derived in the BDZ theory is
\begin{equation}
\displaystyle
I_0^{BDZ} = g(R_c)\, \frac{\lambda_Z}{\sqrt{\pi} R_c^2}\, {\cal V}\ ,
\label{23}
\end{equation}
where one has to evaluate the rate $g(R_{c})$. From the significance
of the current $J$, the drift coefficient $\displaystyle \frac{g}
{k_B T}\, \frac{\partial \Delta F}{\partial R}$ acts as a velocity in
size space \cite{LiPi81}, $d R/\! d t$, which can be determined 
independently by the solution of the macroscopic equations for the 
growth of a droplet. We will consider the diffusion model of droplet
growth which assumes that the droplet of a new phase grows due to the 
diffusion flux of matter through its interface. Then, the problem of 
the liquid$-$gas phase transition is similar to the case of 
precipitation of a substance from a supersaturated solution or 
separation in binary fluids. The only difference is that one has to 
choose the order parameter as the local entropy density in a 
liquid$-$gas system, and as the local concentration in a weekly 
supersaturated solution \cite{Kawa75}. For the latter case the 
spherically symmetric concentration distribution $c(r)$ around a 
droplet of radius $R$ is given by the steady state solution of the 
diffusion equation
\begin{equation}
\displaystyle
D \nabla^2 c(r) = \frac{\partial c(r)}{\partial t} = 0\ ,
\label{23a}
\end{equation}
where $D$ is the solute diffusion coefficient. We obtain
\begin{equation}
\displaystyle
c(r) = c - (c - c_{0R} ) \frac{R}{r}\ ,
\label{23b}
\end{equation}
where 
\begin{equation}
\displaystyle
c_{0R} = c_0 + \frac{R}{r} ( c - c_0 ) 
\label{23c}
\end{equation}
is the equilibrium concentration of the solution at the surface of
a droplet of radius $R$, $c_0$ is the equilibrium concentration of
the solution above the planar surface, and $c$ is the given value 
of the concentration of the supersaturated solution. Substituting
Eq.~(\ref{23c}) in Eq.~(\ref{23b}) and identifying $\delta c$ as
the supersaturation $c - c_0$ in the system, we get
\begin{equation}
\displaystyle
c(r) = c - \delta c \frac{R_c}{R} \left( 1 - \frac{R_c}{R} 
\right)\ .
\label{23d}
\end{equation}
Since the diffusive flux into the droplet $\displaystyle i(r) = D
\frac{\partial c}{\partial r}$ at a droplet surface is equal to the
growth rate of the droplet, one may write
\begin{equation}
\displaystyle
\frac{d R}{d t} = \frac{D\, \delta c}{R}\, \left( 1 - \frac{R_c}{R}
\right)\ .
\label{23e}
\end{equation}
Linearizing this equation around the critical radius and taking into 
account that, by definition of the dynamical prefactor, $R - R_c =
\exp{(-\kappa t)}$, we have finally
\begin{equation}
\displaystyle
\frac{d R}{d t} = |\kappa|\, \left( \frac{R_c}{R} \right)^2\,
(R - R_c)\ ,
\label{24}
\end{equation}
where $\displaystyle |\kappa| = \frac{D\, \delta c}{R^2}$. This 
radial dependence of the droplet growth rate is valid for the 
liquid$-$gas phase transition also. Similar result was obtained 
in Ref.~\cite{LaTu80} by virtue of the hydrodynamic Kotchine 
equations describing the enlargement of a droplet.
Thus, for the critically large droplet we have
\begin{equation}
\displaystyle
g(R_c) = k_B T \left( \frac{d R}{d t} \right)_{R_c}\,
\left[ \frac{\partial \Delta F}{\partial R} \right]_{R_c}^{-1}
= \frac{ |\kappa|}{2 \lambda_Z^2}\, R_c^2\ .
\label{25}
\end{equation}
Substitution of Eq.~(\ref{25}) in Eq.~(\ref{23}) yields
\begin{equation}
\displaystyle
I_0^{BDZ} = \frac{|\kappa|}{2 \sqrt{\pi}}\, \frac{{\cal V}}{\lambda_Z}
\ , \label{26}
\end{equation}
leading again to the result (\ref{22}).

For the sake of simplicity we considered the process where the one
order parameter was sufficient for the description of nucleation.
Obviously, our analysis may be extended to the system of arbitrary
many degrees of freedom. We proposed a direct method allowing one
to link the classical theory to the modern field theory of 
nucleation. It is very important for the comparison that the 
underlying dynamics of the process, that governs the 
cluster growth, should be the same. Trying to compare both theories
Langer and Turski found \cite{LaTu73}, that the numerical predictions
of the critical supersaturation in Xe and CO$_2$ given by the field
nucleation theory are very close to those of the classical theory.
But the rate of molecule accretion on the surface of the critical
cluster, $g (R_c)$, was calculated from simple geometric 
considerations of the kinetic theory of gases. Therefore, the 
analytical expressions of the nucleation rates were different, 
and it was difficult to find conformity between them. 

Recently \cite{KVV95} the Langer theory of nucleation has been 
applied to calculate quark$-$gluon plasma formation in heavy ion 
collisions at the energy of the Brookhaven accelerator. Then the 
nucleation rate derived in the field theory has been compared 
with that of the classical theory. Again it turns out that both 
prefactors are about the same order of magnitude.

The classical nucleation theory is attractive not only because it
predicts nucleation rates in terms of macroscopic quantities. It can
be easily generalized to time-dependent nucleation phenomena. Also,
the classical theory can be extended to the temperature and 
curvature dependences of the interfacial energy, to the deviations of
the shape of the clusters from the ideal spherical shape, etc. In 
spite of these advantages, and although the classical theory appears
to be consistent with the experimental data, it is implied explicitly 
and implicitly that one must consider the classical theory only as a 
crude approximation. This problem becomes clearer now. Using the 
macroscopic diffusion model for the description of a spherically 
growing droplet of liquid we have shown that the classical
nucleation rate can be derived in a form identical to that of the
field nucleation theory. The results obtained in this paper support
the validity of the phenomenological classical theory on the basis 
of the more accurate Langer first-principles theory of nucleation.
A further comparison between these theories may be useful for the 
development of a more rigorous nucleation theory.

{\bf Acknowledgments.}
We would like to thank L. Csernai for stimulating discussions and
suggestions. We wish also to thank A. Vischer for bringing to our 
attention Ref.~\cite{KVV95}. We are grateful to the 
Department of Physics, University of Bergen and to the Institute
for Theoretical Physics, University of Frankfurt for the warm and 
kind hospitality. One of us, L.B., acknowledges support of the 
Alexander von Humboldt Foundation.

\end{document}